\documentclass[preprint2]{aastex}

\usepackage{emulateapj5,apjfonts}
\usepackage{graphicx}
\usepackage{onecolfloat}
\slugcomment{accepted for publication in PASP}

\shorttitle{PMAS: II. PPak}
\shortauthors{Kelz, Verheijen et al.}

\begin{document}

\twocolumn[    % --- for preprint only

\title{PMAS: The Potsdam Multi Aperture Spectrophotometer. \\
       II. The Wide Integral Field Unit PPak}


\author{Andreas Kelz\altaffilmark{1},
        Marc A.W. Verheijen\altaffilmark{1,2},
        Martin M. Roth\altaffilmark{1},\\
        Svend M. Bauer,
        Thomas Becker,
        Jens Paschke,
        Emil Popow,
        Sebastian F. S\'anchez\altaffilmark{3}
}
\affil{Astrophysikalisches Institut Potsdam, An der Sternwarte 16, D-14482 Potsdam, Germany}
\email{akelz@aip.de}
%\and
\author{Uwe Laux\altaffilmark{4}}
\affil{Th\"uringer Landessternwarte Tautenburg, Sternwarte 5, 07778 Tautenburg, Germany}
%\email{laux@tls-tautenburg.de}
%\and

\begin{abstract}

PPak is a new fiber-based Integral Field Unit (IFU), developed at the
Astrophysical Institute Potsdam, implemented as a module into the
existing PMAS spectrograph. The purpose of PPak is to provide both an
extended field-of-view with a large light collecting power for each
spatial element, as well as an adequate spectral resolution. The PPak
system consists of a fiber bundle with 331 object, 36 sky and 15
calibration fibers. The object and sky fibers collect the light from
the focal plane behind a focal reducer lens. The object fibers of PPak,
each 2.7 arcseconds in diameter, provide a contiguous hexagonal
field-of-view of 74 $\times$ 64 arcseconds on the sky, with a filling
factor of 60\%. The operational wavelength range is from 400 to 900~nm.
The PPak-IFU, together with the PMAS spectrograph, are
intended for the study of extended, low surface brightness objects,
offering an optimization of total light-collecting power and spectral
resolution. This paper describes the instrument design, the assembly,
integration and tests, the commissioning and operational procedures,
and presents the measured performance at the telescope.

\end{abstract}

%\keywords{Integral field spectroscopy, 3D spectroscopy, optical design and manufacturing, optical fibers}

]   % --- for preprint only

\altaffiltext{1}{Visiting Astronomer, German-Spanish Astronomical
  Center, Calar Alto, operated by the Max-Planck-Institute for
  Astronomy, Heidelberg, jointly with the Spanish National Commission
  for Astronomy.}
\altaffiltext{2}{Current address: Kapteyn Astronomical Institute,
  Postbus 800, 9700 AV Groningen, The Netherlands}
\altaffiltext{3}{Current address: Centro Astron\'omico Hispano Alem\'an de Calar Alto, C/Jesus Durban Remon 2-2, E-04004 Almeria, Spain}
\altaffiltext{4}{Under contract with AIP.}

%%%%%%%%%%%%%%%%%%%%%%%%%%%%%%%%%%%%%%%%%%%%%%%%%%%%%%%%%%%%%
\section{Introduction}
\label{sect:intro}  % \label{} allows reference to this section

3D-spectrographs (3DS), or Integral Field Units (IFU) exist at many observatories, providing spectra for a large number of spatial elements (``spaxel'') within a 2-dimensional field-of-view,
rather than only along a traditional 1-dimensional spectrograph
slit. Depending on the instrument, up to hundreds or thousands of
spectra are recorded simultaneously in any single exposure. While the
instrumentation suite is diverse and based on various principles of
operation (image slicers, lens-arrays, fiber-bundles or combinations of
these), compromises with respect to field-of-view, spatial sampling,
wavelength coverage, and spectral resolution have to be made, due to
the limited detector space.

Since commissioning in May 2001, the Astrophysical Institute Potsdam
(AIP) successfully operates PMAS, the Potsdam Multi-Aperture
Spectrophotometer, at the Calar Alto 3.5~m Telescope in southern Spain
(Roth et al.\ 2004, Kelz et al.\ 2003a).  An overall description of
the PMAS instrument is given by Roth et al.\ (2005), hereafter
referred to as paper~I.  While PMAS is a unique spectrophotometer,
covering a wide wavelength range from 350~nm to 900~nm, its standard~IFU, a fiber-coupled lens-array, provides 256 spectra and is limited to a maximum integral field-of-view of 16$\times$16 arcseconds on the sky.

Driven by the ``Disk Mass'' project (Verheijen et al.\ 2004), which
requires imaging spectroscopy of nearby face-on galaxies (with typical sizes of 1 arcminute) at intermediate spectral resolution (of R$\ge$8000),
a science case was put forward to develop a larger IFU for PMAS.
Based on the experience with the SparsePak bundle, which was constructed and
commissioned for the 3.5m WIYN telescope at Kitt Peak (Bershady et al.\ 2004, 2005), the PPak (PMAS fiber Package) fiber bundle was designed and built
at the AIP in 2003 as part of the ULTROS project (ultra-deep optical
spectroscopy with PMAS). This new IFU was produced on a short timescale
of approximately six months and with a budget of less than 20.000 Euro
for the hardware components (mainly lenses, filters and fibers).
PPak was successfully integrated within PMAS in December 2003 and commissioned in spring 2004. The PPak-mode of PMAS is now fully operational and routinely employed for the Disk Mass project, as well as for a variety of other
common user programmes, that require large integral-field spectroscopy.

%\onecolumn
\begin{table*}%[ht]
%\begin{deluxetable}{lcccrccrcc}%[ht]
\tabletypesize{\small}
%\tablewidth{0pt}
\caption{Selected IFU instrumental parameters with corresponding spectral capabilities. \label{tab:IFUs}}
%\begin{center}
\begin{tabular}{lcccrccrcc}
%\startdata
\tableline
\tableline
Instrument &     Telescope     &   FoV   &      Spaxel      & Spaxel
 & Filling & Range\tablenotemark{4}& Resol.\tablenotemark{4} &\multicolumn{2}{c}{Grasp}          \\
           &      diameter & (max.)              & size\tablenotemark{1}                  & number\tablenotemark{2}
 & factor\tablenotemark{3}  & ($\lambda_{cov}$)              & ($\lambda/\Delta \lambda$)       &    specific\tablenotemark{5}     &      total\tablenotemark{6}      \\
           &       [m] &   [arcsec]   &     [arcsec]     &
 &         &     [\AA]     &        &[arcsec$^2$m$^2$]&[arcmin$^2$m$^2$]\\
\tableline
PMAS-PPak{$^7$}  & CA     \hfill 3.5 & 74$\times$64 &  2.68 $\oslash$  &
331+36 &   0.60  &     400       &  8000  &  47\phantom{.4} &  4.23
\\
SparsePak  & WIYN   \hfill 3.5 & 72$\times$71 &  4.69 $\oslash$
&75+\phantom{3}7&   0.25  &     260       & 12000  & 138\phantom{.4} &  2.87
       \\
DensePak   & WIYN   \hfill 3.5 & 45$\times$30 &  2.81 $\oslash$
&91+\phantom{3}4&   0.42  &     260       & 20000  &  49\phantom{.4} &  1.25
       \\
INTEGRAL   & WHT    \hfill 4.2 & 34$\times$29 &  2.70 $\oslash$  &
115+20 &   0.67  &     360       &  4200  &  73\phantom{.4} &  2.32
\\
\tableline
VIMOS	   & VLT    \hfill 8.2 & 54$\times$54 & 0.67$\times$0.67 &          6400
&   1.00  &     350       &   220  &  23\phantom{.4} & 40.50\phantom{7} \\
SAURON     & WHT    \hfill 4.2 & 41$\times$33 & 0.94$\times$0.94 &
1577 &   1.00  &     540       &  1250  &  11\phantom{.4} &  4.77           \\
SPIRAL{$^8$}	   & AAT    \hfill 3.9 & 22$\times$11 & 0.7 $\times$0.7  &           512
&   1.00  &     330       &  7600  &   5.4           &  0.77           \\
PMAS-LARR{$^9$}  & CA     \hfill 3.5 & 16$\times$16 & 1.0 $\times$1.0  &
256 &   1.00  &     700       & 6000  &   8.2           &  0.58           \\
OASIS	   & WHT    \hfill 4.2 & 17$\times$12 & 0.42 hex.        &    $\sim$1100
&   1.00  &     370       &  2650  &   2.3           &  0.71           \\
GMOS	   & Gemini~ \hfill 8.1 &  7$\times$5  & 0.2 hex.         &          1000+500 &
  1.00  &     280       &  1700  &   1.8           &  0.49           \\
\tableline
%\enddata \\
\\
\end{tabular}
%\end{center}
\small{$^1$}{~corresponding value to the max. FoV, may depend on fore-optics magnification} \\
\small{$^2$}{~dedicated sky-spaxels are listed separately}\\
\small{$^3$}{~bare fiber bundles (upper part); lens-array-types (lower part)} \\
\small{$^4$}{~selected values only; may span wide range depending on configuration and wavelength} \\
\small{$^5$}{~specific grasp = telescope area [m$^2$] $\times$ spaxel size [arcsec$^2$]} \\
\small{$^6$}{~total grasp = telescope area [m$^2$] $\times$ spaxel size [arcsec$^2$] $\times$ number of spaxels } \\
\small{$^7$}{~PPak-IFU and PMAS spectrograph with 2nd order gratings} \\
\small{$^8$}{~SPIRAL with Littrow spectrograph (de-commissioned)} \\
\small{$^9$}{~Lens-array IFU and PMAS spectrograph with 1st order gratings} \\
%\end{deluxetable}
\end{table*}
%\twocolumn

As a bare fiber-bundle IFU, PPak is based on earlier developments,
such as DensePak (Barden \& Wade 1988), INTEGRAL (Arribas et al.\ 1998) and SparsePak (Bershady et al.\ 2004).
Similar to the above instruments, PPak opts for rather large fibers,
that can not properly sample the (seeing-limited) image,
but collect more flux and allow for wide fields.
PPak and SparsePak span 74$\times$64 and 72$\times$71 arcseconds on the sky
respectively and provide the largest fields-of-view of any IFU available
worldwide (see Table~\ref{tab:IFUs}).
Additionally, a single PPak fiber with 5.7~arcsec$^2$ on the sky, collects twice the amount of light at the 3.5m telescope, than a single spatial element of the VIMOS-IFU (LeFevre et al.\ 2003) at the 8.2m-VLT (see specific grasp in Table~\ref{tab:IFUs}).
PPak is attached to the efficient PMAS fiber-spectrograph,
that provides resolutions R=$\lambda/\Delta \lambda$ from 800 to 8000,
or corresponding spectral powers (defined as the product of resolution
times the number of spectral resolution elements N$_{\Delta \lambda}$ $\lambda/\Delta \lambda$) between 10$^6$ to 10$^7$.
The PPak fibers and optics are optimized for wavelengths between 400--900~nm.

\begin{figure*}%[h]
\epsscale{2.0}
\plotone{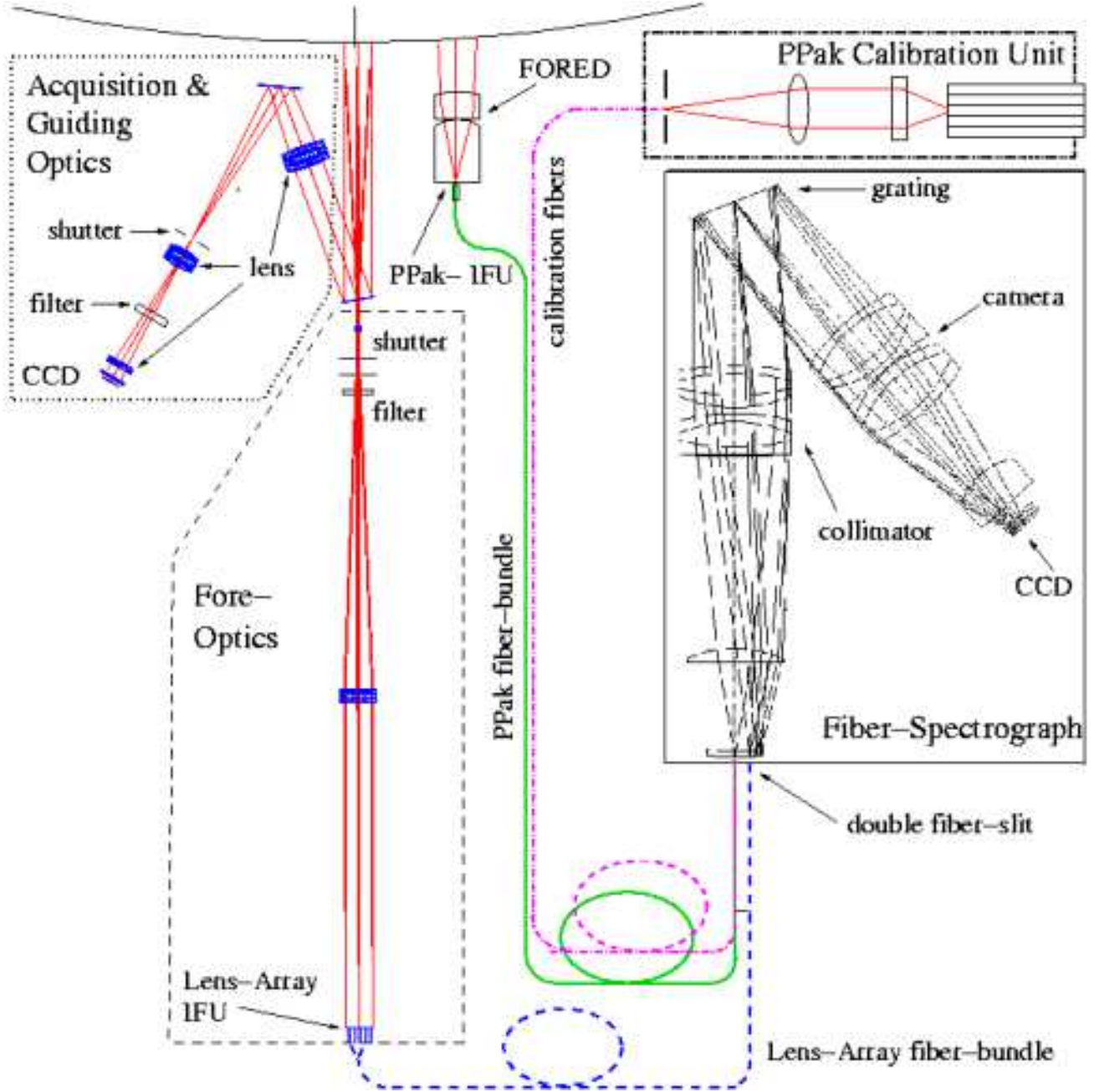}
\caption{The principle of operation for the two PMAS IFUs
	 at the Cassegrain focal station.
         The LARR-IFU is on-axis and consists of fore-optics,
	 a lens-array (LARR) and a fiber-bundle (dashed lines).
	 The PPak-IFU is located off-axis and features a
	 focal reducer lens (FORED) and a bare fiber-bundle.
	 A dedicated PPak calibration unit can illuminate
	 additional fibers (dashed-dotted lines).
	Both bundles connect to one spectrograph (solid outline).
 	 The acquisition and guiding system (dotted outline)
	 can be used by both modes
	(see text for further explanation).}
\label{fig:ppak_principle}
\epsscale{1.0}
\end{figure*}

Therefore, PPak is ideally suited for
spectroscopic studies of extended astronomical objects with low
surface brightness, such as the outskirts of spiral galaxies, where
sufficient signal is more important than detailed spatial
resolution.
The PMAS+PPak configuration offers a powerful combination of light-collecting power or grasp, wavelength range, and spectral resolution. \\

This paper is organized as follows: \S~\ref{sect:design} presents the
instrument and its opto-mechanical design. \S~\ref{sect:mai}
summarizes the manufacture, assembly, and integration of the PPak
components.  \S~\ref{sect:operations} describes the operational
procedure during observation, the data reduction, and visualization
tools. The instrument performance at the telescope and test results
are given in \S~\ref{sect:performance}.

%%%%%%%%%%%%%%%%%%%%%%%%%%%%%%%%%%%%%%%%%%%%%%%%%%%%%%%%%%%%%

%%%%%%%%%%%%%%%%%%%%%%%%%%%%%%%%%%%%%%%%%%%%%%%%%%%%%%%%%%%%%
\section{Instrument Description}
\label{sect:design}

The baseline parameters for the PPak development were, to provide a contiguous sampled field-of-view of at least 1~arcminute across with high specific grasp per spaxel and adequate resolution at the spectrograph.
PPak needed to be designed as an unforeseen add-on module to the
existing, Cassegrain mounted, PMAS instrument. Therefore, certain
space constraints dictated the overall design.
Likewise, the PMAS spectrograph hardware and performance had to be taken as given.
The existing PMAS grating set (see table~2 of paper~I) includes gratings
with 300, 600 and 1200 l/mm, of which two can be used in the 2nd spectral order.
Taking advantage of anamorphic demagnification (see section \ref{sect:2ndorder}), spectral resolutions of R$\sim$8000 can be achieved with the I1200 and J1200 gratings, if the width of the pseudo-slit of the PMAS
spectrograph does not exceed 150~microns, which respectively limited the fiber core diameters. Around 400 of these fibers can be accommodated at the spectrograph slit  with acceptable separations and cross-talk.
The PMAS spectrograph accepts a F/3 beam, which implies that, allowing for some focal ratio degradation, the fibers can be fed with up to F/3.3 in the focal plane, which sets the plate scale and fiber grasp.
The purchase of even higher-line density or holographic reflection gratings
was no consideration at the time of the PPak development.

Fig.~\ref{fig:ppak_principle} illustrates the principle of operation
for both the pre-existing lens-array IFU and the new PPak IFU: For the
lens-array-mode, the telescope focal plane image is magnified by a fore-optics (dashed box)~(Roth et al.\ 2003), then spatially sampled by a
lens-array and re-configured by an optical fiber module (Kelz et
al.\ 2003b).

PPak, on the other hand, is equipped with a focal
reducer lens (FORED) in front of the telescope's focal plane, which maximizes the field-of-view, and provides the required plate scale and f-number. The
subsequent fiber bundle (solid line) by-passes the fore-optics and lens-array
and bridges the distance of around 3~meters to the spectrograph.
Additional fibers (dash-dotted) connect the spectrograph to a dedicated  calibration unit. At the spectrograph entrance, the fibers from both the lens-array and the PPak-IFU form two parallel slits, of which only one is
active during observing. The PPak-IFU is placed 6 arcminutes off-axis,
as not to obstruct the existing field for the direct imaging
camera. This allows to use the A\&G optics (dotted box) for target
acquisition and guiding in both the PPak- and lens-array-IFU mode. The various
components are described in more detail in the following subsections.

%\pagebreak

%%-----------------------------------------------------------
\subsection{Focal Reducer Lens}
\label{sect:FORED_design}

\begin{figure*}[ht!]
\centerline{
\epsscale{1.7}
\plotone{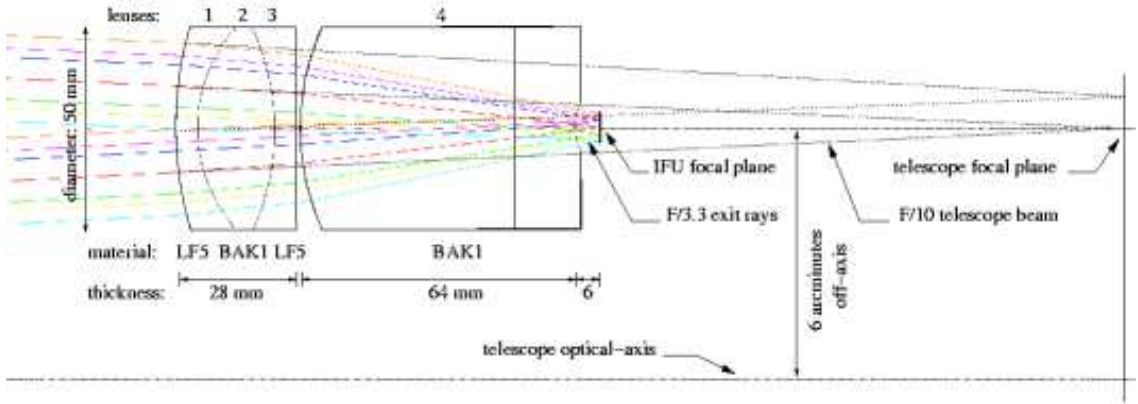}}
\caption{Ray tracing of the focal reducer system. The original
  telescope focal plane with its F/10 rays is shown to the right (dotted lines).
  The focal reducer converts the rays to F/3.3 (solid lines) and creates a focal
  plane 6~mm after the last lens. (optical design by U.L.)}
\label{fig:fored_ray}
\end{figure*}
\begin{figure*} [ht!]
\plotone{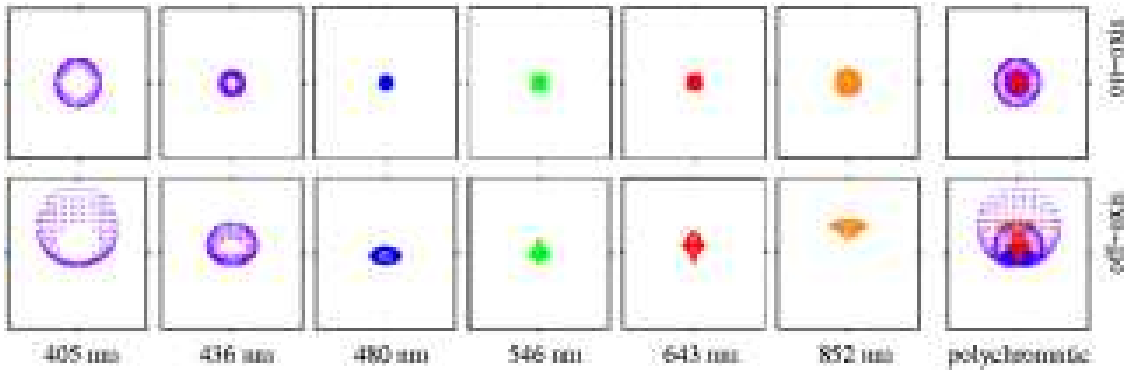}
\caption{Spot diagrams at the focal plane of the focal reducer system at design
  wavelengths of: 405, 436, 480, 546, 643, 852~nm, and for a polychromatic spot. The top and bottom rows represent on-axis and 3.5~mm off-axis spots respectively.
The box size is 100~$\mu$m, corresponding to 1.8~arcseconds, which is smaller than the diameter of a fiber core.}
\label{fig:spotdiagrams}
\epsscale{1.0}
\end{figure*}

The focal reducer lens (FORED) immediately in front of the fiber
bundle reduces the focal length of the Calar Alto 3.5m telescope from
35000~mm to 11550~mm, changes the plate-scale from $5''.89$ per mm to
$17''.85$ per mm, and converts the telescope F-number from F/10 to
F/3.3. The focal reducer is a 4-lens system, consisting of a triplet
and a thick singlet lens (see Fig.~\ref{fig:fored_ray}),
which creates the required telecentric exit rays.
The system has four glass-to-air interfaces,
which are treated with anti-reflective coatings. The individual lenses
of the triplet are made of LF5, BaK1, LF5, respectively, while the
fourth lens is made of BaK1. The thickness of the triplet is 28~mm, that of the single lens 64~mm, while their diameters are 50~mm.
The lenses are optimized for a
wavelength interval between 400 and 850 nm, i.e. neglecting the blue
part of the spectrum that otherwise is accessible with the PMAS
spectrograph.

The image quality, that the focal reducer provides, was optimized to match the fiber size of 150 $\mu$m and does not deteriorate the
point-spread-function beyond the typical seeing (see spot diagrams in
Fig.~\ref{fig:spotdiagrams}).
As the light is coupled to optical fibers, telecentric exit rays are
required. The focal reducer lens creates a telecentric field of 7 mm (=125 arcsec) in
diameter. Additionally, the system can be placed up to 65~mm off-axis,
which is necessary due to space constraints within the existing PMAS
instrument. The off-axis position of the focal reducer causes a small
telecentric offset of 0.25~degree.  While in principle this offset
could be compensated completely with a tilt correction of the entire
fiber array, it was considered a small and negligible
change in terms of the incoming F-number
(see Fig.~12 of Bershady et al. 2004).

\begin{figure}[ht!]
\begin{center}
\plotone{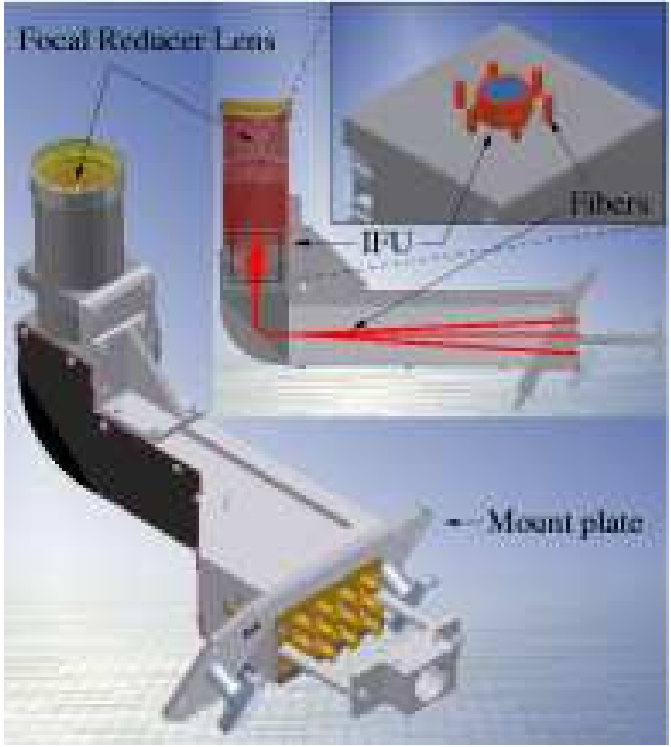}
\caption{Three CAD views of the mount, that holds the focal reducer lenses,
  the fiber head and optional filters. Due to space constraints, the
  fibers are bend by 90$^\circ$ below the IFU and exit sideways.
  A flexure analysis of the mount stability vs. inclination, yields a maximum  deformation of 4~$\mu$m at the top part with respect to the fixed mount plate.
  (mechanical design by S.M.B.)}
\vspace{5mm}
\label{fig:CAD_FORED_mount}
\end{center}
\end{figure}

A common, compact and stiff mount for the focal reducer and the PPak-IFU was
designed in order to make sure, that the fiber bundle is firmly
attached to the focal reducer image plane, which is located 6~mm
behind the lens (see Fig.~\ref{fig:CAD_FORED_mount}).  This mount
also allows the insertion of bandpass or interference filters (with diameters
of 50~mm or 2~inch) in front of the lens.  Any mechanical flexure due
to a changing gravity vector was calculated to be 4~$\mu$m ($\approx$
0$''$.1) in the worst case.  In the spatial direction, this is much smaller than the fiber sampling size (of 150~$\mu$m) or effects caused by seeing
(0$''$.5 $\approx$ 30~$\mu$m). In terms of focal accuracy, this amount
of flexure is negligible. Note, that due to the limited pointing accuracy
of the telescope, it is practically impossible to measure flexure effects
of this magnitude, if present at all.

%%-----------------------------------------------------------
\subsection{PPak IFU}

\begin{figure}[h]
\plotone{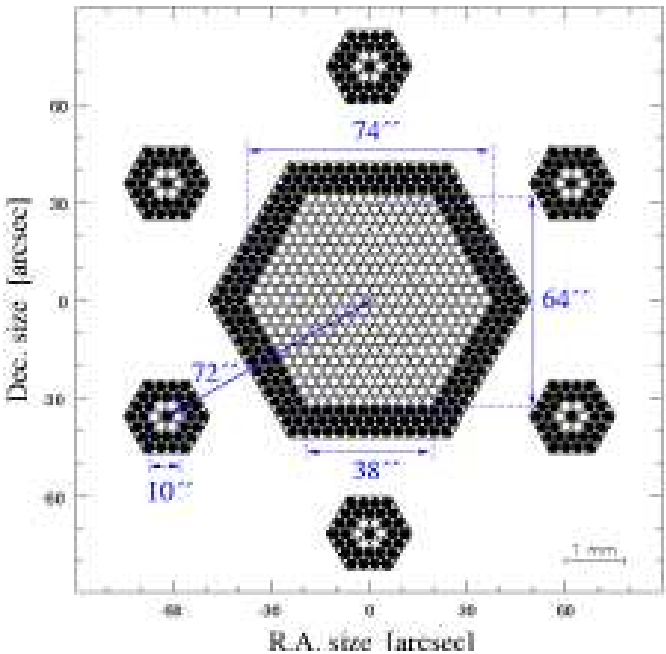}
\caption{Layout and dimensions of the PPak-IFU. The central hexagonal
  is made up of 331 object fibers, surrounded by six sky-IFUs. Note,
  that only the white circles represent active fibers, while the black
  ones are protective buffers. Each circle represents the combined fiber core, cladding and buffer material. While the physical size of the central IFU is
  just 4 mm, its coverage on the sky is more than 1 minute of arc.}
\label{fig:PPak_size}
\end{figure}

The final PPak design features 331 fibers in a densest packed hexagonal grid
with a maximum diameter of 74$^{\prime\prime}$, while each fiber
projects to 2$^{\prime\prime}$.68 in diameter on the sky. The fiber-to-fiber
pitch is 3.6 arcseconds. The projected fiber area with 5.7 arcsec$^2$ is comparable to DensePak or INTEGRAL, but smaller than for SparsePak. However, the larger number of fibers allows the observer much freedom, to apply adaptive binning of spaxels to increase the signal-to-noise.

Additional 36 fibers are distributed amongst
six `mini-IFUs' and placed 72$^{\prime\prime}$ away from the center to
sample the surrounding sky (see Fig.~\ref{fig:PPak_size}). Not shown
are 15 extra fibers, that are not part of the IFU, but are connected to
a calibration unit. These calibration fibers can be illuminated with light from spectral line lamps during the science exposures. This provides a synchronous spectral calibration and keeps track of any image shifts at the spectrograph detector (see section 4.9 of paper~I).

A fair amount of additional fibers (shown in black) is placed
around the active (white) fibers for protective buffer purposes and
to avoid increased FRD edge-effects (see Figs.~ 8 and 9 in Bershady et al., 2004) of the outer science fibers. These
buffer fibers have a length of $\sim$ 70~mm and terminate inside the mount, just below the IFU head. Table~\ref{tab:PPak_fibers} gives a summary of the total fiber
breakdown.

\begin{table}[ht]
\caption{Breakdown of total number of fibers for PPak.}
\begin{center}
\begin{tabular}{rrrr}
\tableline
\tableline
\# Fibers    & $\circ$ Active & $\bullet$ Buffer & $\circ+\bullet$ Total \\
\tableline
object		& 331		& 216		& 547 	\\
sky		&  36 		& 186		& 222	\\
calibration	&  15 		&  22 		&  37 	\\
\tableline
total		& 382		& 424		& 806 	\\
\tableline
\end{tabular}
\end{center}
\label{tab:PPak_fibers}
\end{table}

\begin{figure}[h]
\plotone{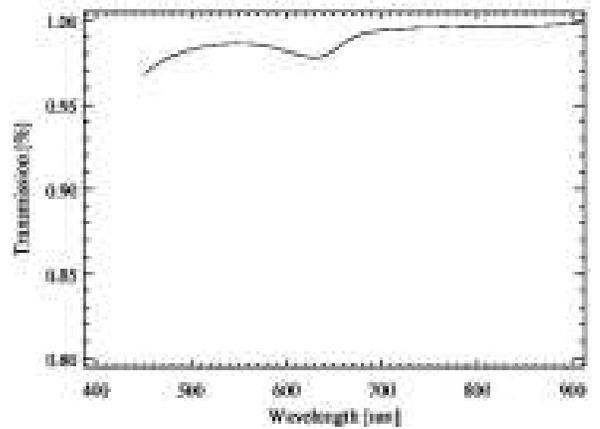}
\caption{Internal fiber transmission plot for 4m of FIP fiber as supplied by the manufacturer Polymicro Technologies LLC. PPak uses 3.5m long fibers of this series, with a NA=0.22, a core diameter of 150$\mu$m and a core/cladding ratio of 1:1.1.
No additional transmission measurements were done within this project.
Note, that end losses or losses from FRD effects are not included in the above data.}
\label{fig:fiber-trans}
\end{figure}

%%-----------------------------------------------------------
\subsection{Fiber Slit}

The output ends of the fibers are placed side-by-side to form a long
fiber-slit. For practical reasons and to assist data
reduction, the overall fiber-slit is divided into 12 blocks (called
slitlets). A slitlet is 7.5~mm wide and features 32 v-grooves with a
spacing of 0.234~mm (see Fig.~\ref{fig:fibslit}, right).
%The distance between the edges of the fiber cores is 84 microns.
Given a spectrograph magnification of 0.6 and 15~$\mu$m pixel,
a fiber core projects to 6 pixel at the detector, with a pitch of 9.4 pixel.
The chosen spacing is a trade-off to minimize cross-talk and the overall slit-length, because of edge vignetting.

Contrary to other IFUs, that are add-on units to existing spectrographs, PMAS features a designated fiber-spectrograph (see paper~I). The spectrograph
collimator is a f=450~mm, F/3 system, and therefore the optics accepts
the whole fiber output cone without the need of additional
beam-converting micro-lenses.
The spectrograph optics require a curved fiber-slit,
that directly couples to the first lens of the collimator.
The fibers are mounted parallel to each other, but
terminate at different lengths, to form a curved focal surface (see Fig.~\ref{fig:fibslit}, left).

%%-----------------------------------------------------------
\subsection{Slit to Sky Mapping}

\begin{figure*}[ht!]
\epsscale{1.5}
\plotone{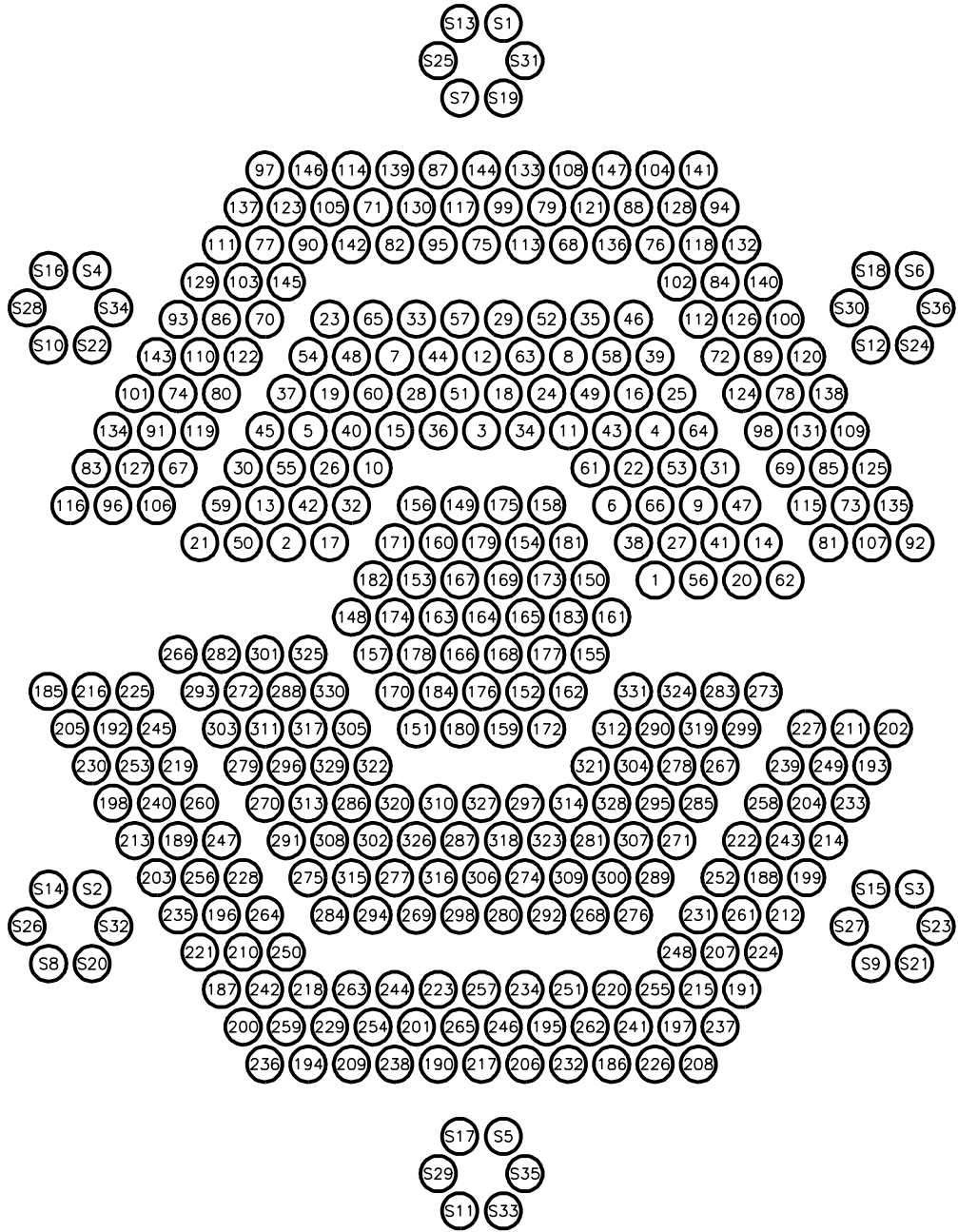}
\caption{In the focal plane, five main segments can be identified
  which map to distinct regions along the slit (see
  Tab.~\ref{tab:PPak_slit}). North is up and east is left. The orientation is fixed on the sky as the instrument can not be rotated.}
\label{fig:PPak_layout}
\epsscale{1.0}
\end{figure*}

The focal plane geometry of the PPak IFU is largely determined by the Disk Mass project.
The arrangement of the fibers in the focal plane and how they map to
the slit follows a quasi-random fashion. The central IFU can be
divided in five main segments (see Fig.~\ref{fig:PPak_layout}). The
central segment, with fibers numbered 148 to 184, maps to the central
part of the slit. The fibers in the two intermediate segments of the
IFU (1 to 66 for the northern part, and 266 to 331 for the southern
part), map to the edges of the slit. The fibers in the two outer IFU
segments (67 to 147 for the northern part, and 185 to 265 for the
southern part), terminate halfway between the center and the edges of the slit
(see Tab.~\ref{tab:PPak_slit}).
A reason for this arrangement was, that for typical targets, such as galaxies,
the surface brightness falls off rapidly away from the center.
As the outer fibers carry the weaker signals, any additional vignetting or aberration effects, predominantly at the edge of the slit, shall be avoided.
Care was taken, that fibers which are adjacent in the focal plane, are well separated in the slit.
The aim was to minimize any systematic effects, that purely depend on the location of fibers within the spectrograph.
Note, that this is an opposite philosophy to other instrument layouts,
such as SPIRAL (Lee \& Taylor
2000), where adjacent fibers at the sky remain adjacent at the slit.

Also note, that the central row at the IFU, starting in the east with fiber number 185 and ending in the west with number 92, contains fibers from all 5 segments. This implies, that drifting a star across the central row produces 21 spectra distributed over the entire CCD, nearly sampling the full range of optical paths
through the spectrograph.

Each slitlet carries 3 sky fibers. Those 3 sky fibers are
distributed in three non-adjacent sky-IFUs. In this way, each
triplet of sky fibers on a slitlet spans a triangular area on the sky,
which surrounds the main IFU. In other words, the sky is sampled symmetrically around the object and is well distributed within the slit, again to avoid any instrumental biases.
This scheme results in 36 sky-fibers altogther, which is twice the `optimum' number of dedicated sky fibers as calculated by Bershady et al. (2004, Fig.3), but helps to limit systematic errors in the sky subtraction.

\begin{table}[h]
\setlength{\tabcolsep}{1mm}
\caption{The location of object ({1--331}), sky ({\bf S1--S36}), \& calib\-ration ({\bf C1--C15}) fibers and gaps ($\times$) along the slit (1,1$\rightarrow$12,32). Underlined numbers indicate  a new IFU segment.}
{\footnotesize
\begin{center}
\begin{tabular}{c|cccccccccccc}
\tableline
\tableline
Groove~~~&\multicolumn{12}{c}{Slitlet number} \\
number~~~&    1   &     2   &     3   &     4   &    5   &     6   &     7   &     8   &     9   &    10   &    11   &    12   \\
\tableline
  1 &{\bf  C1}&{\bf  C3}&{\bf  C4}&{\bf  C5}&{\bf  C6}&{\bf  C7}&{\bf  C9}&{\bf C10}&{\bf C11}&{\bf C12}&{\bf C13}&{\bf C14}\\
  2 &{\bf  C2}&    27   &    55   &    83   &   111   &   139   &   166   &   194   &   222   &   250   &   278   &   306   \\
  3 & $\times$&    28   &    56   &    84   &   112   &   140   &   167   &   195   &   223   &   251   &   279   &   307   \\
  4 &       1 &    29   &    57   &    85   &   113   &   141   &   168   &   196   &   224   &   252   &   280   &   308   \\
  5 &       2 &    30   &    58   &    86   &   114   &   142   &   169   &   197   &   225   &   253   &   281   &   309   \\
  6 &{\bf  S1}&{\bf  S4}&{\bf  S7}&{\bf S10}&{\bf S13}&{\bf S16}&{\bf S19}&{\bf S22}&{\bf S25}&{\bf S28}&{\bf S31}&{\bf S34}\\
  7 &       3 &    31   &    59   &    87   &   115   &   143   &   170   &   198   &   226   &   254   &   282   &   310   \\
  8 &       4 &    32   &    60   &    88   &   116   &   144   &   171   &   199   &   227   &   255   &   283   &   311   \\
  9 &       5 &    33   &    61   &    89   &   117   &   145   &   172   &   200   &   228   &   256   &   284   &   312   \\
 10 &       6 &    34   &    62   &    90   &   118   &   146   &   173   &   201   &   229   &   257   &   285   &   313   \\
 11 &       7 &    35   &    63   &    91   &   119   &\underline{147}&174&   202   &   230   &   258   &   286   &   314   \\
 12 &       8 &    36   &    64   &    92   &   120   &   148   &   175   &   203   &   231   &   259   &   287   &   315   \\
 13 &       9 &    37   &    65   &    93   &   121   &   149   &   176   &   204   &   232   &   260   &   288   &   316   \\
 14 &      10 &    38   &\underline{66}& 94 &   122   &   150   &   177   &   205   &   233   &   261   &   289   &   317   \\
 15 &      11 &    39   &    67   &    95   &   123   &   151   &   178   &   206   &   234   &   262   &   290   &   318   \\
 16 &      12 &    40   &    68   &    96   &   124   &   152   &   179   &   207   &   235   &   263   &   291   &   319   \\
 17 &{\bf S2}&{\bf  S5}&{\bf  S8}&{\bf S11}&{\bf S14}&{\bf S17}&{\bf S20}&{\bf S23}&{\bf S26}&{\bf S29}&{\bf S32}&{\bf S36}\\
 18 &      13 &    41   &    69   &    97   &   125   &   153   &   180   &   208   &   236   &   264   &   292   &   320   \\
 19 &      14 &    42   &    70   &    98   &   126   &   154   &   181   &   209   &   237   &\underline{265}&293&   321   \\
 20 &      15 &    43   &    71   &    99   &   127   &   155   &   182   &   210   &   238   &   266   &   294   &   322   \\
 21 &      16 &    44   &    72   &   100   &   128   &   156   &   183   &   211   &   239   &   267   &   295   &   323   \\
 22 &      17 &    45   &    73   &   101   &   129   &   157   &\underline{184}&212&   240   &   268   &   296   &   324   \\
 23 &      18 &    46   &    74   &   102   &   130   &   158   &   185   &   213   &   241   &   269   &   297   &   325   \\
 24 &      19 &    47   &    75   &   103   &   131   &   159   &   186   &   214   &   242   &   270   &   298   &   326   \\
 25 &      20 &    48   &    76   &   104   &   132   &   160   &   187   &   215   &   243   &   271   &   299   &   327   \\
 26 &      21 &    49   &    77   &   105   &   133   &   161   &   188   &   216   &   244   &   272   &   300   &   328   \\
 27 &      22 &    50   &    78   &   106   &   134   &   162   &   189   &   217   &   245   &   273   &   301   &   329   \\
 28 &{\bf  S3}&{\bf  S6}&{\bf  S9}&{\bf S12}&{\bf S15}&{\bf S18}&{\bf S21}&{\bf S24}&{\bf S27}&{\bf S30}&{\bf S33}&{\bf S36}\\
 29 &      23 &    51   &    79   &   107   &   135   &   163   &   190   &   218   &   246   &   274   &   302   &   330   \\
 30 &      24 &    52   &    80   &   108   &   136   &   164   &   191   &   219   &   247   &   275   &   303   &   331   \\
 31 &      25 &    53   &    81   &   109   &   137   &   165   &   192   &   220   &   248   &   276   &   304   & $\times$\\
 32 &      26 &    54   &    82   &   110   &   138   &{\bf  C8}&   193   &   221   &   249   &   277   &   305   &{\bf C15}\\
\tableline
\end{tabular}
\end{center}
}
\label{tab:PPak_slit}
\end{table}

%%-----------------------------------------------------------
\subsection{Fiber Loop Box}

Stress on optical fibers increases the focal ratio degradation (FRD)
with consequences for the overall optical system performance
(see Barden 1998, Parry \& Carrasco 1990, Ramsey 1988, Schmoll et
al.\ 2003, and references therein).
To minimize FRD, and therefore the loss of information, the fibers are
inserted into protective, friction-free 3-layer furcation tubings, and only bent with tolerable radii. Roughly 2~m behind the IFU and some 40~cm in front of the fiber-slit, the protective tubing is interrupted, to
allow the fibers to form a loop. The fiber loops are placed
inside an enclosed box of 30 $\times$ 30~cm, where they are being kept
in groups of 32 and placed into separate sections, divided by Teflon sheets (see Fig.~\ref{fig:loopbox}).
The loop box serves two functions:
Any pull on a fiber results in a change of the individual loop diameter,
which avoids the fibers from being torn. Secondly, the loops provide a
reservoir of extra fiber length, which is needed during assembly and integration.
Note, that both the IFU and the spectrograph are mounted at the Cassegrain station and remain fixed with respect to each other.
Therefore, no stress-relief cabling, as for bench-mounted instruments with long ($>$10m) fiber length is required.

%%---------------------------------------------------------
\subsection{PPak Calibration Unit}

\begin{center}
\begin{figure}[ht!]
\plotone{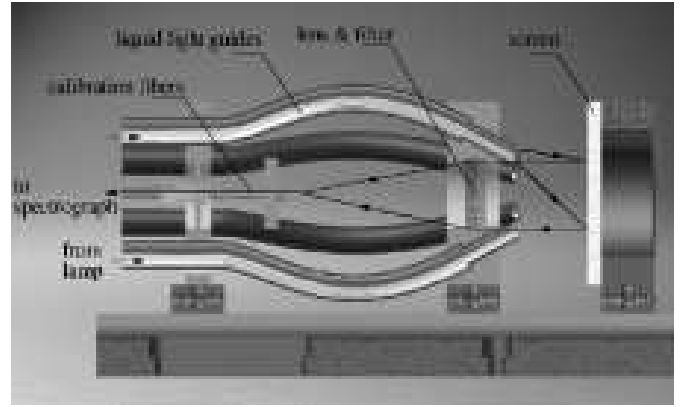}
\caption{The PPak Calibration Unit: Liquid-light-guides illuminate a
  white diffuser screen, from which the light reflects backwards. A
  relay lens couples the light with the correct F-number into the calibration fibers. An optional filter can be inserted. Up to 5 light-guides offer the
  possibility to combine the light from different lamps.}
\label{fig:PPCU}
\end{figure}
\end{center}

Fifteen fibers that are distributed along the fiber-slit are diverted
from the rest of the fiber-bundle, as their input ends are not placed
at the telescope focal plane but connected to the PPak Calibration
Unit (PPCU). This unit is made from standard {\em OWIS} laboratory
equipment and consists of five liquid light guides ({\em Lumatec},
Germany), a white diffuser screen, a relay lens and the calibration
fibers themselves (see Fig.~\ref{fig:PPCU}).  Four liquid light guides
illuminate the diffuser screen with light from the various calibration
lamps (such as Halogen continuum, Mercury, Neon and Thorium/Argon).
A fraction of the light is picked-up by the relay lens and
focused onto the calibration fibers, with the same F-number as for the
science fibers, and with an object distance at infinity. As the lamps are
placed within electronic boxes and feature individual shutters,
any combination of calibration light (and respective exposure times)
can be fed into the calibration unit. The calibration fibers can be illuminated separately from or
simultaneously with the object fibers, allowing the observer
flexibility with respect to calibration strategy and needs.  Finally,
a change of lamps or a swap to a spare lamp is easily done by
re-connecting the light guide(s), without any changes to the
calibration unit itself.

%%%%%%%%%%%%%%%%%%%%%%%%%%%%%%%%%%%%%%%%%%%%%%%%%%%%%%%%%%%%%
\section{Manufacture, assembly and integration}
\label{sect:mai}

%%-----------------------------------------------------------
\subsection{Lens Optics}

The fabrication of the focal reducer lens was contracted
to {\em Pr\"azisionsoptik Gera}, (Germany), based on specifications
from the optical design calculations by U.L. (see
\S~\ref{sect:FORED_design}).  These calculations were repeated as soon
as the glasses had been procured from {\em Schott}, (Germany), and the
index of refraction had been measured at the design wavelengths in
order to optimize the design. %(Fig.~\ref{fig:FORED_POG})
Note, that due to its thickness, the fourth lens was made from two pieces.
The individual lenses were cemented using optical compound K57, produced by {\em Carl Zeiss Jena}.
All glass--air surfaces were treated with a Balzers broad-band AR coating,
yielding a transmission of $>$98\% across the design wavelength range.

The first lens of the spectrograph collimator (diameter D=100~mm, curvature R=218mm), to which the fiber-slit connects, was produced by
{\em Carl Zeiss Jena} (Germany).

%%-----------------------------------------------------------
\subsection{Fiber Cables}

After tests in the AIP laboratories, silica/silica, step-index fibers with core/cladding/buffer diameters of 150/165/195 microns, low OH core and NA=0.22 of the series FIP150165195 from {\em Polymicro Technologies Inc.} (USA) were selected
(see Fig.~\ref{fig:fiber-trans}). The fiber
bundle was manufactured as follows: Firstly, the fibers were cut to a
length of 3.5~m and one end was polished manually (smallest grain
size=0.3 $\mu$m). The reason for polishing the exit surfaces at this early stage was, that the assembled fiber-slit is curved and therefore difficult to polish. At the input end, the fibers were assembled into the IFU head first and polished afterwards.
Copying parts of the SPIRAL-B design (Lee \& Taylor
2000), a three-layer polypropylene-KEVLAR-PVC furcation tube from {\em
  Northern Lights Cable} (USA) was cut to lengths and fitted with
connector screws on both ends. Altogether 367 fibers were inserted
into 12 protective tubings, each tube carrying the object and sky
fibers from one slitlet. The 15 calibration fibers were put into an additional tube.

%%---------------------------------------------------------
\subsection{Fiber-Slit Assembly}

\begin{center}
\begin{figure}[h]
\plotone{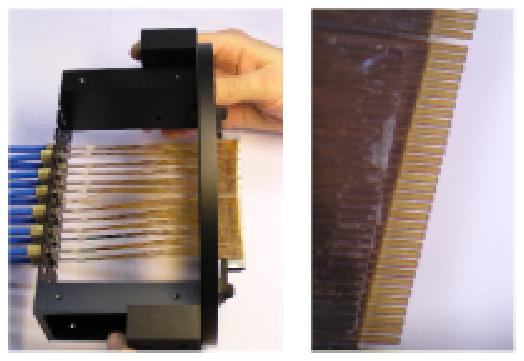}
\caption{Left: photo of the fiber-slit unit with all 382 fibers and 13 cables
  in place. Note the overall curvature of the slit. Right: a magnified
  view of an individual slitlet (out of twelve) with 32~fibers (width = 7.6~mm,
  fiber-spacing = 0.234~mm)}
\label{fig:fibslit}
\end{figure}
\end{center}
\vspace{-7mm}

The polished ends of the fibers were glued onto the 12 slitlets, using
EPO-TEK 301-2 non-shrinking two-component epoxy from {\em Polytek}.
Each slitlet typically holds 28 object fibers, 3 sky fibers and 1
calibration fiber (see Tab.~\ref{tab:PPak_slit} for details). The
outermost calibration fibers ({\bf C1} and {\bf C2} at one end, and
{\bf C15} at the other end of the slit) are separated from the object
fibers by one empty groove. Three sky fibers (from 3 different sky-IFUs) are distributed uniformly amongst the 28 object fibers.
To ensure a correct and repetitive fiber alignment, an assembly jig
was produced, that holds both the fibers and the slitlet in place and allows
an accurate end termination of each fiber against a dummy surface with the correct curvature. After the correct alignment was controlled visually,
the fibers were clamped temporarily and then glued onto the slitlet block.
Given, that the v-grooves are 100 times longer than a fiber diameter, and  manufacturing tolerances of $<$0.01~mm were achieved, the alignment and the end positioning of the fibers was done to within a few microns accuracy.
Altogether, twelve slitlets, each
carrying 31 or 32 fibers each, were mounted side-by-side on a common
stage, creating a fiber-slit of 94~mm length in total.
The slitlets can be moved and locked individually.
In this way, any length variations between slitlets are irrelevant, as each slitlet can be brought forward, until the fibers touch the collimator lens surface.
The overall unit includes mounts for the cables, the fiber-slit and
the collimator lens as well as protective covers (Fig.~\ref{fig:fibslit}, left).

The fibers were repeatedly cleaned with methanol and water in an ultra-sonic bath prior and after the assembly, to remove any contamination from the surfaces. The quality was controlled by inspections of the fiber ends using a video microscope. Illumination of individual fibers yielded their
positions within each slitlet, and the fibers were labeled according
to a pre-defined position table (see Fig.~\ref{fig:PPak_layout}).

%%---------------------------------------------------------
\subsection{Fiber-Head Assembly}

\begin{center}
\begin{figure}[h!]
\plotone{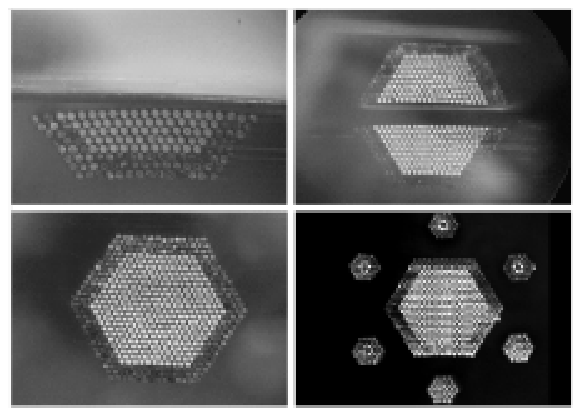} % for 1-col
\caption{Top and left: The assembly of the IFU was done row-by-row for each
  half separately and then merged into a single mount.
  Bottom right: Image of the PPak fiber head taken with a $\times 20$
  magnification to determine the exact fiber positions.
  The surrounding rows of darker fibers are in-active,
  but protect and buffer the central science and sky fibers.}
\label{fig:IFUbuilt}
\end{figure}
\end{center}

At the input (i.e. the IFU) side, the individually labeled object and buffer
fibers were ordered and pre-assembled row-by-row on a piece of sticky
tape. No additional glue was applied at this stage. A mount was
manufactured at the AIP workshop, that features a central hexagonal
opening, with precision steps to aid the correct fiber alignment.
The milling precision was on the order of 1/100~mm (=5\% of a fiber diameter),
which was more than adequate to ensure, that the fibers form a dense-packed arrangement.
The main IFU was built up by inserting each row of
fibers (27 rows in total), into the hexagonal mount, with
the fibers extending $\approx$5~mm beyond the mount surface. For
practical reasons, this was done separately for the two halves (see
Fig.~\ref{fig:IFUbuilt}), which were put together and locked mechanically thereafter.
In fact, not the precision of the mount, but the added tolerance of the sizes of the overall number of fibers limited the correct alignment. The maximum deviation from a regular hexagonal grid occurs for one outermost row and was measured to be of the order of 10\% of a fiber diameter, corresponding to a misalignment of approximately 0.3~ arcsec.

\vspace{-8mm}

\begin{center}
\begin{figure}[h]
\plotone{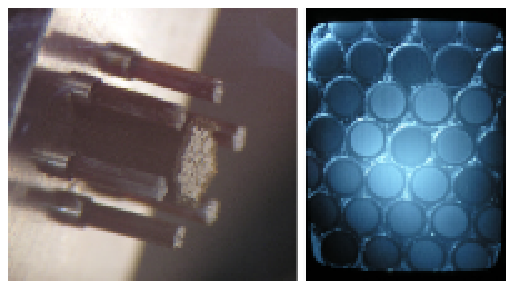}
\caption{Left: Magnified view of the assembled IFU (including
  sky-IFUs) before polishing. Right: a video inspection image with $\times$ 8 magnification of the polished fibers.}
\label{fig:IFUpolish}
\end{figure}
\end{center}

Six additional circular drillings around the central opening accommodate
the sky-fiber IFUs. The assembly of the sky-IFUs worked similar with respect to the row-by-row approach. Each sky-IFU consists of 6 sky and 31 short
buffer fibers (Fig.~\ref{fig:PPak_size}), which form a densest pack of 7 fibers across. This was inserted into a steel ferule with a matching inner diameter of 1.4~mm. Subsequently, the ferules were glued into the drillings of
the mount (Fig.~\ref{fig:IFUpolish}, left).
After assembly, the IFU-mount was pointed downward and the
extending fibers were immersed in a bath of epoxy which worked its way
upward in-between the fibers and through the mount by means of
capillary force. In this way, the fibers are glued together and to the
metal mount without introducing additional stress.
After the epoxy had cured, the entire fiber head, including the six sky-IFUs, was polished using a custom-made polishing stage (as described in Kelz et al.\ 2004).  Polishing sheets from {\em Newport} and {\em Data
Optics}, featuring grain sizes from 30 to 0.3~$\mu$m were used. The
surface quality was inspected regularly during the polishing process using a video microscope with 4--16 times magnification (Fig.~\ref{fig:IFUpolish}, right). This allowed the projection of highly magnified fiber images onto a monitor. In conjunction with a variety of viewing angles and illuminations,
scratches at the order of 1~micron were visible and could be polished out.
The end requirements were, to have no obvious surface defects, such as partial breakages of core or cladding material, no scratches larger than 1\% of the fiber diameter ($\leq$1.5~$\mu$m), and a visually flat and perpendicular endface.

\vspace{-8mm}

%%---------------------------------------------------------
\subsection{Integration of PPak into PMAS}

\begin{center}
\begin{figure}[h]
\plotone{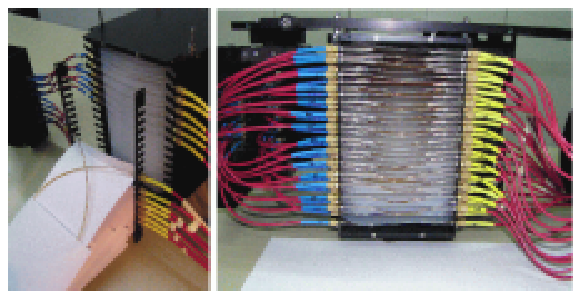}
\caption{Left: photo of the fiber-loop box, showing how a section of bare fibers is looped and inserted into a compartment. Right: photo of the fully filled, but still open loop box. The cables to the left connect to the slit, the ones to the right to the IFUs. 16 cables contain the 256 LARR-fibers, 12 cables the 367 PPak-fibers, one cable the 15 calibration fibers.}
\label{fig:loopbox}
\end{figure}
\end{center}

Optical gel (code 0406, n=1.46) from {\em Cargille Laboratory} was
applied between the spectrograph collimator lens and the fiber-slit(s)
to match the refractive indices.  The loop-box was filled and
closed (see Fig.~\ref{fig:loopbox}), and all 13 PPak-cables inserted into a common flexure tube.
Inspections revealed that no fibers were broken or damaged during the process of manufacture, assembly and integration.

\begin{center}
\begin{figure}[h]
\plotone{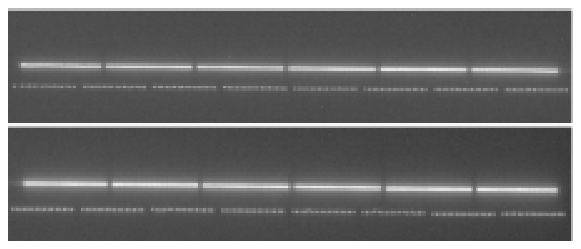}
\caption{Zero order image of the double fiber-slit. The fainter rows with 16
  blocks belong to the 16 $\times$ 16 lens-array-IFU, the brighter rows
  with 12 blocks belong to PPak. For better clarity, the
  slit (physical size=94mm) is split in two, the top image connects to
  the left of the bottom one.}
\label{fig:DIFU}
\end{figure}
\end{center}

During the construction and initial commissioning, the PPak bundle was an entity
from end to end. This implied, that the existing lens-array fiber-module needed to be dis-mounted from the PMAS instrument, to make space for the PPak-module.
As this is a time-consuming and potentially hazardous undertaking, both fiber modules, i.e. the fiber-slits and loop-boxes, were merged into a single unit
(called double-IFU) in October 2004. The double-slit consists of two parallel
rows, featuring the 256 lens-array fibers and the 382 PPak fibers (see
Fig.~\ref{fig:DIFU}). The spacing between the slits is approximately
2~mm. The fore-optics and the two IFUs remain physically separate
units.  While both IFUs can not be used simultaneously, it is easy and safe to
change the configuration during daytime.  A change of modes
between the lens-array-IFU and the PPak-IFU involves a hardware switch to
select a different shutter, to reconnect the internal lamps to the
respective calibration units, and to cover the IFU
that is not in use.

%%%%%%%%%%%%%%%%%%%%%%%%%%%%%%%%%%%%%%%%%%%%%%%%%%%%%%%%%%%%%
\section{Operations}
\label{sect:operations}

%%-----------------------------------------------------------
\subsection{Calibration}

While the entire lens-array-IFU can be illuminated from a deployable
internal calibration unit, the position of the PPak-IFU, being
off-axis and in front of the telescope focal plane, is outside the
opto-mechanical range of the  original calibration unit.
Instead, PPak calibrations must be performed with dome or sky flatfield exposures.  Flatfield exposures with external light sources (continuum, arc lamps) yield the fiber-to-fiber responses, the wavelength calibration, and the
position information required for the purpose of accurately tracing
and subsequently extracting the spectra from a recorded image. These
calibration images should be obtained at least once per night and for
each grating setting.

\begin{center}
\begin{figure}[h]
\plotone{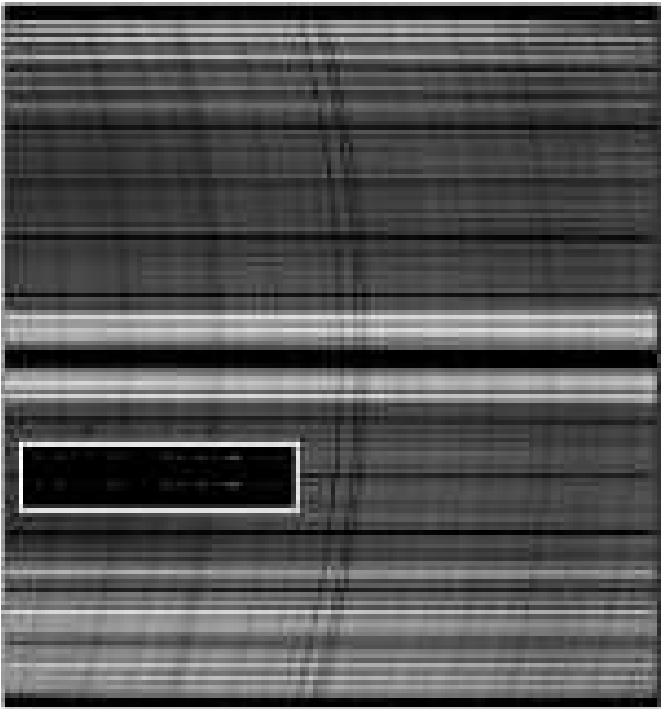}
\caption{Raw CCD frame illustrating the basic features of the PPak data format:
382 spectra, grouped in 12 blocks (vertical axis) versus wavelength (horizontal axis). A central gap is provided for the use with a mosaic-CCD. The dark arcs are absorption lines (not wavelength calibrated). The bright dots are emission line spots in the calibration spectra. (see insert and Fig.~\ref{fig:PPak_anamag} for a magnified area near the center of the CCD).
The distribution of light within the slit illustrates the mapping philosophy (compare with Fig.~\ref{fig:PPak_layout} and Table~\ref{tab:PPak_slit}).}
\label{fig:rawframe}
\end{figure}
\end{center}

The best spectrograph focus is found by illuminating the calibration
fibers only, using the internal spectral line lamps.  This will yield
well separated emission spots across the entire CCD chip.  Preferable,
the calibration fibers should always be illuminated by a spectral line
lamp while a science exposure is taken. This allows the tracing of
any image shifts or spectrograph de-focus during the data reduction
for each science frame (see Fig.~\ref{fig:rawframe}).

%%-----------------------------------------------------------
\subsection{Instrument Control Software}

The AG-OPTICS system for acquisition and guiding has been described in
paper~I.  As opposed to the lens-array-IFU, which is positioned at the center of the A\&G field-of-view (of 3$'$.2 $\times$ 3$'$.2),
the PPak-IFU is located off-axis,
295$''$ to the South and 206$''$ to the West.  Since the offset is
known to the instrument control software, the basic functionality for
field acquisition and guiding remains unchanged. The A\&G instrument
control software (pics\_ag) has an option to over-plot the PPak outline
or even the position of the 331 fibers to a freshly obtained
acquisition frame and allows for an offset pointing to center the
object of interest onto the PPak-IFU. The position of a guiding box
around a guide-star on the A\&G frame can be stored and recalled. This
allows for the accurate re-positioning of a guide-star on subsequent
nights to within 0.2 arcseconds. Note, that this procedure has been successfully applied using the finer sampling lens-array IFU, which, due to
its distance, is more likely subject to relative flexure, than the closely mounted PPak unit (see Fig.~\ref{fig:ppak_principle}).

The IDL-based PMAS Instrument Control Software (PICS) includes some
additional features for PPak operations, namely the option to include
calibration light within the science frames (e.g. 5 $\times$
10~seconds of ThAr distributed equally within an overall exposure time
of 30~minutes, say) and to center, offset or mosaic-point the
PPak-IFU.

Note, that the nod-and-shuffle (or beam switching) mode, that is
available for the lens-array-IFU, is possible with PPak too, but at the expense
of a higher cross-talk, as the spacing between the PPak spectra on the detector
is smaller.

%%-----------------------------------------------------------
\subsection{Data Reduction Software}

\begin{figure}[h]
\plotone{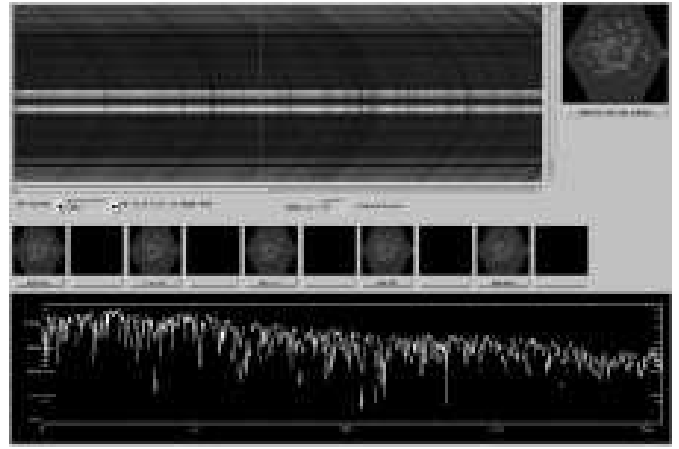} % for 1-col
\caption{Graphical User Interface (GUI) of the PPak-online data
  reduction software, written in IDL by T.B. The display contains the (not
  yet calibrated) stacked spectra (top panel), re-constructed maps at
  various wavelength cuts (central panels) and a spectrum of one
  selected spaxel (bottom panel).}
\label{fig:PPak_online}
\end{figure}

Based on the ``P3d'' data reduction software package (Becker 2002, see
paper I), an adapted version for PPak (PPAK\_online) was written by
T.B.  The program can be used for a quick-look inspection of the data
quality and for a reconstruction of maps, while observing at the
telescope.
The code is written in IDL and allows to process the raw
data and to eliminate the specific instrumental signature.  The
subroutines include: bias and dark subtraction, cosmic cleaning,
spectra tracing, flexure compensation, spectra extraction,
flat-fielding, and wavelength calibration.  The full P3d package also
allows a CCD pixel-to-pixel response variation, stray-light modeling,
and wavelength-dependent fiber response calibration. There are also
various custom-made IDL utilities for the visualization of stacked
spectra, maps, and individual (or co-added) spectra (see
Fig.~\ref{fig:PPak_online}).  These utilities are available within
GUIs and from the IDL command line, supporting the use of scripts.  It
is also possible to call the E3D visualization tool from within
PPAK\_online, providing further features, such as interpolated maps,
line fitting, etc (see S\'anchez 2004, S\'anchez et al.\ 2004).  Also,
PPak data was successfully reduced, using the hydra package within
IRAF.

%%%%%%%%%%%%%%%%%%%%%%%%%%%%%%%%%%%%%%%%%%%%%%%%%%%%%%%%%%%%%
\section{Performance}
\label{sect:performance}

%%-----------------------------------------------------------
\subsection{Throughput}

The instrumental throughput was obtained using two methods.  Firstly,
the observed flux of spectrophotometric standard stars was compared to
tabulated values from the literature.  Secondly, the relative
throughput of domeflat exposures for the lens-array-IFU and the PPak-IFU was
determined.  The reason for the second approach is, that often the
actual atmospheric conditions at Calar Alto are either non-photometric
or not known well enough, as to determine the true instrumental
response unambiguously.  However, the instrumental efficiency using
the lens-array-IFU was well established previously (see paper~I), and
therefore a relative response measurement can yield the
PPak-throughput.

\begin{figure}[h]
\plotone{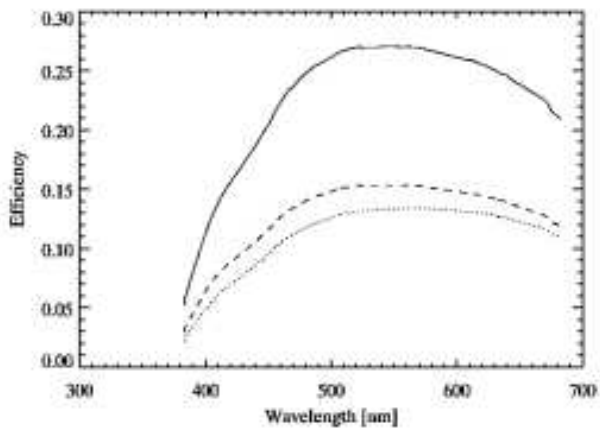}
\caption{PMAS+PPak efficiency, measured from a single-fiber standard
  star integration.  Plotted are the total throughput (dotted curve),
  the efficiency of telescope and instrument combined (dashed curve),
  and the pure instrumental efficiency (solid curve), taking the
  atmospheric extinction and the telescope reflectivity into account.
  Note: the total response curve is based on telescope
  transmission under less than optimal conditions (dust).}
\label{fig:PPak_effi1}
\end{figure}

Fig.~\ref{fig:PPak_effi1} plots the directly measured PMAS+PPak
efficiency.  The lower (dotted) curve gives the total throughput, from
top of the atmosphere to the detector.  It was obtained by comparing
the flux of the spectrophotometric standard star BD~+75~325, observed
on Nov. 20, 2004 at an airmass of 1.3, to the expected flux as given
by Oke (1990).  This total efficiency $\eta$ includes the instrument
$\eta_{ins}$, the atmosphere $\eta_{atm}$, and the telescope
$\eta_{tel}$. The middle (dashed) curve represents the efficiency from
top of the telescope to the detector, i.e. taking atmospheric
extinction and airmass into account.  The atmospheric extinction
coefficients for each wavelength were calculated by scaling typical
extinction tables for Calar Alto (Hopp \& Fernandez 2001) to
$k_{ext}$=0.14 mag and $\eta_{atm}$=0.85, which was the measured
extinction in the V-band during the night.  The top (solid) curve is
the pure instrumental throughput $\eta_{ins}$, from the telescope
focal plane to the detector.
The instrumental configuration included the PPak-IFU without any filters
and the spectrograph with the V300 grating in 1st order (300~l/mm, blaze angle=4.3$^\circ$, $\alpha$=16$^\circ$, $\lambda_{cen}$=542~nm).  The
throughput of the primary mirror was derived from reflectivity
measurements obtained routinely at Calar
Alto\footnote{\url{http://dbserv.caha.es/iris/index.asp}}. % ake3-9
Assuming a similar value of 75\% for the secondary reflectivity, the
telescope efficiency was estimated to be $\eta_{tel}$=0.57 in V at the
time of observation.

Note, that the plots in Fig.~\ref{fig:PPak_effi1} present {\em lower
  limits} in that the flux lost outside the finite aperture of a
single fiber was neglected.  Applying an aperture correction (along
the lines of CCD aperture photometry techniques, e.g.\ Howell 1989)
based on the measured seeing FWHM of 1$''$.1 in V, and assuming a
Gaussian approximation to the point-spread-function, we estimate a
correction factor of 1.15, i.e.\ a peak efficiency of 31\%.

This value agrees reasonably well with the comparison of domeflat
exposures, taken with both the lens-array and the PPak-IFU, while the
spectrograph setup, using a V300 grating, remained unchanged.
The PPak configuration was found to have a 1.5 times higher throughput,
than the lens-array-IFU, resulting in a peak efficiency of 30\% rather than 20\%, respectively.
As the coupling of both fiber-slits towards the spectrograph is similar and
the length differences between the lens-array and PPak-fibers is only 1.5~m,
we attribute this difference in efficiency to the fore-optics and input coupling in particular. As shown in \S~4.3 of paper~I, the lens-array IFU
throughput suffers from two effects: firstly from light losses caused by the extended parts of the lens-array PSFs (i.e. stray light and diffraction spikes) and secondly from the median mis-alignment between the micro-pupil images and the fiber cores. Both these problems are not present in the PPak-design, so that a bare fiber-bundle fed by a large fore-optics lens
is more efficient.

Instrumental throughput estimates for other PMAS gratings were
bootstrapped from the lens-array-IFU efficiency data as shown in Fig.~15 of
paper~I. Using a grating blazed in R (600~l/mm, blaze angle=13.9$^\circ$, $\lambda$=530--810~nm), the instrumental efficiency peaks at
36\% between 600--700~nm. Note, that the PPak configuration has not
been optimized in the blue and that both the image quality of the focal reducer lens,
and the transmission of the fibers and the lenses rapidly
decrease below 400~nm. The groove density of the gratings (e.g. 300,
600 or 1200 l/mm) only has a minor effect on the efficiency. Those
gratings, which are in use in 2nd order (I1200, J1200), show a
significantly lower efficiency due to intrinsic grating effects and a
geometrical overfill of the grating at large tilt angles
(see~\S~\ref{sect:2ndorder}).

%%-----------------------------------------------------------
\subsection{Fiber Response and Cross-talk}

\begin{figure}[h]
\includegraphics[width=0.35\textwidth,angle=90]{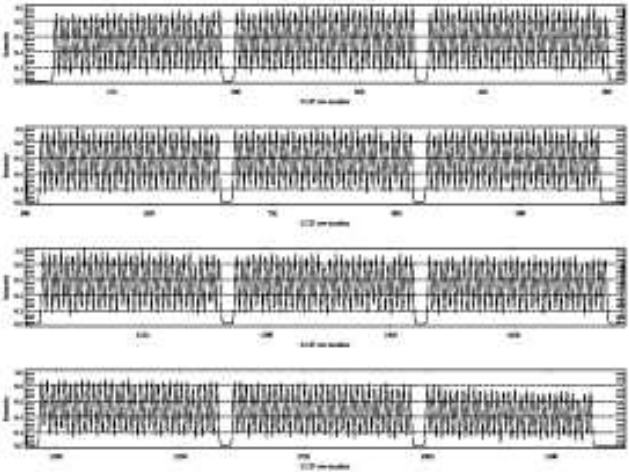} % for 1-col
\caption{A cross-dispersion cut through a raw `domeflat' exposure
  featuring all 367 spectra, that belong to the PPak-IFU (the
  calibration fibers were not illuminated).  Note the gaps between
  the twelve slitlet blocks. }
\label{fig:PPak_cdc1}
\end{figure}

\begin{figure}[h]
\includegraphics[width=0.35\textwidth,angle=90]{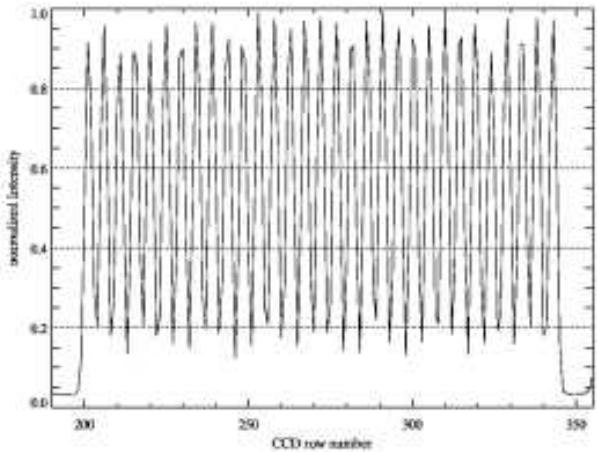} % for 1-col
\caption{Typical fiber-to-fiber throughput variation.
         Zoom of Fig.~\ref{fig:PPak_cdc1}: Shown is one
  slitlet with 31 spectra, which are well separated down to 20\% of
  the peak value.}
\label{fig:PPak_cdc2}
\end{figure}

Visual inspection of the either forward or backward illuminated fiber
bundle yielded a apparent uniform fiber response across the slit and
the IFU, respectively. This was confirmed by sky- and domeflat
exposures taken at the telescope.  Fig.~\ref{fig:PPak_cdc1} shows a
cross-dispersion cut, roughly at 500~nm, through a raw domeflat
exposure with the illuminated 331 `object' and 36 `sky' fibers. The
vignetting of the spectrograph optics towards the edges and the degree of flatness of the fiber-to-fiber response can be judged from this plot. Note, that there is an overall slope in the intensity level, which is believed to result from two effects: the slit is not perfectly centered on the optical axis of the spectrograph and the undersized flatfield screen in the dome, does in fact not provide a uniform and flat illumination across the field. Fig.~\ref{fig:PPak_cdc2} is a
zoomed version of the previous, showing a central slitlet only. The
maxima of the normalized intensities range between 0.89 and 0.98.  In
zero-th order, the fiber core projects to 6~pixels on the CCD.  The
pitch between individual fibers is 9.4~pixels,
resulting in moderate cross-talk and a typical inter-order minimum
intensity of 20\% of the peak level.

The final amount of cross-talk does not only depend on the instrumental performance, but also on the actual data reduction. A profile-fitting extraction method is capable to allocate the overlapping wings of the flux distribution towards the individual spectra. In this way, cross-talk can be dis-entangled better, than using an extraction scheme with fixed pixel numbers (see Becker,\ 2002).
Apart from spectra extraction, other parameters,
such as the option to use on-chip binning in the spatial direction,
or the (sagittal and tangential) focus setting of the spectrograph, will result in different levels of cross-talk. If these are of  concern and what level is acceptable, will depend on the particular science case.

\subsection{Scattered Light}

Scattered light effects are mainly depending on the selected grating,
wavelength range and order, i.e. the grating position. In particular,
ghost images were noticed at certain second order wavelength settings,
where the grating is overfilled and its mount is highly inclined
towards the incoming beam. The exact origin of these ghost images is
subject of further investigation.

\begin{center}
\begin{figure}[h]
\plotone{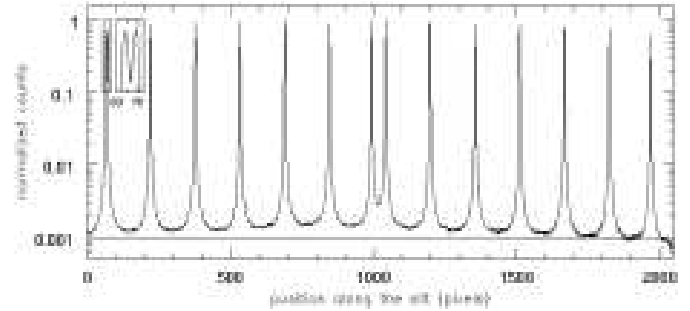}
\caption{Normalized average of 100 columns in the cross-dispersion
  direction near the center of the CCD. Only the 15 calibration
  fibers are illuminated with halogen light, while the 367 IFU fibers are dark. This allows the measurement of the extent of the wings of the flux distribution to low light levels.}
\label{fig:ScatLight}
\end{figure}
\end{center}

In addition to ghost images, most complex optical systems show extended
wings of the point-spread-function at low intensity levels, which are generally
attributed to ``scattered light", although the precise physical origin of this
``halo" around the PSF core is difficult to assess.
The relatively large separation of the PPak calibration fibers presented us
with the opportunity to obtain high signal-to-noise cross-dispersion
profiles from well-exposed continuum lamp calibration exposures, and map
the scattered light level as illustrated in Fig.~\ref{fig:ScatLight}.
The overlapping individual scattered light halos from these calibration
spectra generate a level of $\approx$0.1\% of the peak counts in the areas
in between these spectra.
Despite the fact, that these light levels are low, they can contaminate
the neighboring spectra, depending on the exact method of spectra extraction.
Using a box-like extraction, with a width of 10 fiber-pitches to both sides, around a single, illuminated spectrum, the contribution from the integrated scattered light, can be on the order of 8--14\%.
This value highly depends on the assumed width of a spectrum, the position of the spectrum on the chip and the wavelength.

%\vspace{-5mm}

Becker (2002) has implemented in the P3d data reduction package
a profile-fitting extraction method, which is based on empirically
determined cross-dispersion profiles as a function of wavelength for
each spectrum, which is, however, presently only available for data
obtained with the PMAS lens-array IFU. The iterative scheme of this
code is capable of (1) measuring and eliminating the crosstalk between
adjacent spectra (on the order of 0.5\% for the lens-array), and
(2) simultaneously solving for a model of diffuse scattered light over
the face of the detector (less than 1\% of the average peak intensity).
While this level was considered negligible for normal PMAS data,
the application of the scattered light model to data from the MPFS
instrument at the Selentchuk 6m Telescope, which does have significant
stray-light patterns, proved to be essential to eliminate systematic
errors (Becker 2002). Implementing a profile fitting extraction
routine for PPak data, as well as a thorough characterization of
cross-talk and stray-light properties for the various grating setups
at different grating tilts is a goal with a future upgraded version
of the P3d software.

%%-----------------------------------------------------------
\subsection{Usage in the Second Spectral Order}
\label{sect:2ndorder}

\begin{figure}[h]
\plotone{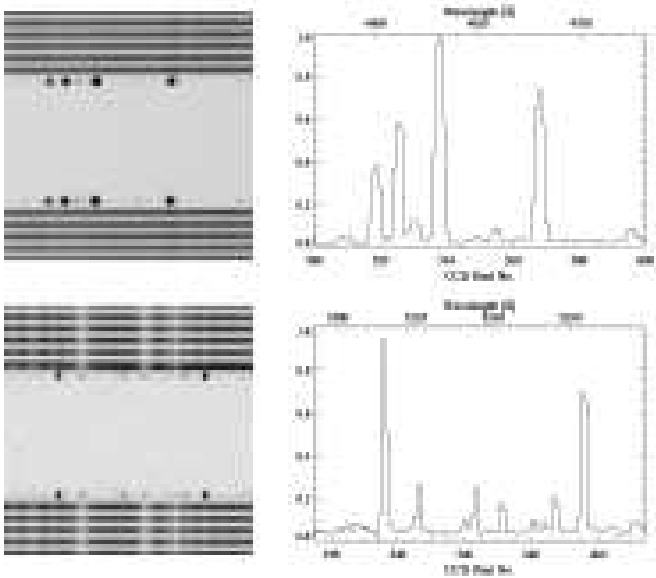} % for 1-column
\caption{Anamorphic magnification. Top: V600 1st order flatfield
exposure with two ThAr calibration spectra from a 100$\times$100
binned pixel region near the the center of the CCD (binning factor
2$\times$2).  Bottom: J1200 in 2nd order (backward), same binning
factor.
Both plots are shown with a negative greyscale stretch. As a result
of anamorphic demagnification, the 2nd order emission line spots are
significantly compressed in the (horizontal) direction of dispersion,
compared with the perfectly round appearance in first order.
The anamorphic factors are r=1.08, and r=0.49, respectively. The spectra
to the right show the corresponding ThAr emission line profiles for
the same region. Note that anamorphic magnification acts only in the
direction of dispersion. For a more detailed explanation, see text.}
\label{fig:PPak_anamag}
\end{figure}

PMAS was initially designed and built as a spectrophotometer for
low/medium spectral resolution, with the goal to maximize wavelength
coverage rather than spectral resolution. More recently, the main
science driver for retrofitting PMAS with the PPak-IFU demanded
R$\approx$8000 (Verheijen et al.\ 2004), which could not be satisfied
with the current fiber size (i.e. pseudo-slit-width) and any of the
standard gratings in 1st order. A test with two gratings in 2nd order
(I1200: blaze angle=37$^\circ$, 1200~l/mm and J1200: blaze angle=46$^\circ$, 1200~l/mm), however, yielded satisfactory results in the wavelength region near 520~nm. Fig.~\ref{fig:PPak_anamag}
illustrates that not only the roughly two-fold increase of linear
dispersion, but also the effect of anamorphic demagnification helps to
improve the spectral resolution (e.g.\ Schweizer 1979). Considering
the basic grating equation: $sin(\alpha) + sin(\beta) = n g \lambda$,
where $\alpha$ and $\beta$ are the angles between the grating normal
and the incident and diffracted beams, respectively, $n$ is the
spectral order, $g$ the groove density in [l/mm], and $\lambda$ the
wavelength, the anamorphic (de)magnification $r$ of the slit width is
given by: $r = {cos(\alpha)}/{cos(\beta)}$.

For a typical setup in first order, $r$ is close to unity, and the
effect is often negligible. In second order, however, the grating tilt
is more extreme so that $r$ becomes significantly different from
1. The upper panel of Fig.~\ref{fig:PPak_anamag} shows a small central
region of a raw CCD frame from a sky flatfield exposure which was
taken on Nov.~8, 2004, using the V600 grating in 1st order with a
grating tilt of $\alpha$=15$^\circ$.  The anamorphic magnification of
this setup is $r$=1.08.
Therefore, the two rows of calibration spectra with ThAr emission
lines present a close to perfect round spot appearance. The
corresponding plot with emission line profiles reveals a FWHM of
$\approx$3 pixels (from a 2x2 binned CCD frame), matching exactly the
expected width of 6 unbinned pixels.

The lower panel shows the same situation for a setup with the J1200
grating in second order, mounted in the ``{\em backward}''
orientation (see paper~I), i.e.\ the grating normal facing the camera, and a grating
tilt of $\alpha$=63$^\circ$.  This sky flatfield exposure was obtained
on May~4, 2005.  In contrast to the V600 exposure from above, the
emission line spots are now significantly compressed in the dispersion
direction, and the line profiles are extremely sharp with a FWHM of
$\approx$1.5 pixels (2x binned).  The anamorphic magnification is
$r$=0.49. The resolving power obtained in this configuration with the
150~$\mu$m PPak fibers is R$\approx$7900.  Note, that in order to
maintain sharp images and this relatively high spectral resolution,
careful focusing, restriction of exposure times, and the avoidance of hour angles with adverse flexure effects (see paper I) are of utmost importance.

The usage of the high-resolution mode comes at the expense of a lower
throughput. Table~\ref{GRATEFF} lists the relative efficiencies $T_{rel}$ (in ADU/s/\AA) of the
I1200 and J1200 gratings in second order, forward (fwd) and backward (bwd)
oriented, respectively, with reference to the V1200 grating in first
order.

\begin{table}[ht]
\label{GRATEFF}
\caption{Relative grating throughput $T_{rel}$ and (de)magnification $r$ at $\lambda$=515~nm}
\begin{center}
\begin{tabular}{lrcrcr}
\tableline
\tableline
Grating     & Blaze[$^{\circ}$] & $n$ & $\alpha$[$^{\circ}$] & $r$ & $T_{rel}[\%]$   \\
\tableline
V600 (fwd)    & 8.6   & 1st   & 11.5  & 1.14 	& 100  \\
V1200 (fwd)   & 17.5  & 1st   & 1.7   & 1.31 	& 100  \\
V1200 (bwd)   & 17.5  & 1st   & 40.3  & 0.76 	& 95   \\
I1200 (fwd)   & 36.8  & 2nd   & 20.5  & 2.03    & 40   \\
I1200 (bwd)   & 36.8  & 2nd   & 62.5  & 0.49    & 55   \\
J1200 (fwd)   & 46.0  & 2nd   & 20.5  & 2.03    & 52   \\
J1200 (bwd)   & 46.0  & 2nd   & 62.5  & 0.49    & 70   \\
\tableline
\end{tabular}
\end{center}
\end{table}

%%%%%%%%%%%%%%%%%%%%%%%%%%%%%%%%%%%%%%%%%%%%%%%%%%%%%%%%%%%%%
\section{Summary}
\label{sect:summary}

A new Integral-Field-Unit, based on the fiber-bundle technique, providing  high grasp and a large field was developed and successfully commissioned for the existing PMAS 3D-instrument.
The central PPak-IFU features 331 object-fibers, which, projected by the 3.5~m Calar Alto telescope, span a hexagonal field-of-view of 74$\times$64~
arcseconds with a filling factor of 60\%. The individual spaxel (fiber) size is 2$^{\prime\prime}$.7 across, yielding a total grasp of 15200~arcsec$^2$m$^2$ at this telescope.
An additional 36 fibers are distributed over six sky-IFUs, which surround the main IFU at a distance of 72$^{\prime\prime}$ from the field center, allowing a good coverage and subtraction of the sky background. For calibration purposes, 15 fibers can be illuminated independently with arc lamps during a science exposure, and can  keep track of spectral resolution and image shifts.
A summary of the technical parameters is given in
Table~\ref{tab:pmas_param}. Further details regarding the PMAS
spectrograph, available gratings and filters are given in paper~I, or
can be found online\footnote{\url{http://www.caha.es/pmas}}.

\begin{figure*}[ht!]
\begin{center}
\includegraphics[width=0.3\textwidth,angle=-90]{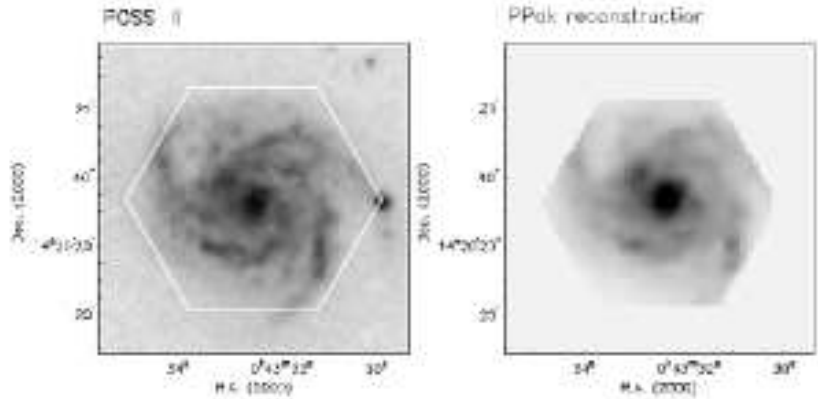} % for 1-column
\caption{Comparison between the POSS-II R-band image (left panel)
and the PPAK reconstructed image of the galaxy UGC 463 (right panel).
The PPAK data were obtained using the V300 grating, centered at $\approx$5300\AA. Once reduced, the reconstructed image was created using E3D
(S\'anchez,\ 2004): a 3D cube was created adopting a natural neighbor interpolation
scheme to a common grid of 1$''$.35/pixel. After that, a 2D image was produced
by co-adding the flux in the wavelength range between 4500 and 6000~\AA.}
\label{fig:UGC463}
\end{center}
\end{figure*}

The combination of spaxels with high grasp and the PMAS
spectrograph with high efficiency and wide wavelength coverage, makes PPak a powerful tool for the study of extended low-surface
brightness objects, which require a high light collecting power and a large field-of-view.
Fig.~\ref{fig:UGC463}, gives an example of the galaxy UGC~463, that was observed for the Disk Mass project.
Despite the rather crude sampling of the fibers, the basic morphological structures of the galaxy seen in the POSS-II image (spiral arms, stars clumps...), are clearly visible in the PPAK reconstructed image.
Apart from the ability to create mono- and polychromatic images from the
resulting data, one exposure with PPak yields 331 spatially resolved spectra  of the target. The high number of fibers at the outer and fainter parts
of the galaxy, offers the observer the option to adaptively bin spaxels as to increase the signal-to-noise further.

During 2004, PPak was available on a shared-risk basis, as its usage involved a complex change-over between the lens-array and PPak fibers,
which needed to be done by AIP staff.
With the integration of the two parallel fiber-slits, the PPak-IFU was
permanently installed and is offered as common-user instrument from 2005 onwards.
Two further upgrades are scheduled: a new mount to ease the
exchange of order separating filters, and adjustments to the
PPak data reduction software.
Since commissioning, PPak attracted considerable interest from several observers, with the result, that 9 observing runs with a total of 26 nights, plus 9 `service buffer A' nights, were granted to PMAS-PPAK within its first year by the Calar Alto TAC. In 2005, approximately 50\% of the allocated time for PMAS is used with the PPak-IFU. Throughout these runs, the PPak module worked without failure.
\begin{table}%%[ht]
\caption{Summary of main PPak + PMAS instrumental parameters.}
\begin{center}
\begin{tabular}{lr}
\tableline
\tableline
{\bf PPak-IFU:} & \\
\tableline
principle design	& focal reducer + fiber-bundle \\
focal reducer lens	& F/10 to F/3.3, dia=50~mm \\
plate-scale 		& $17''.85~$/mm \\
fiber configuration	& 331 object, 36 sky, 15 calibration \\
PPak - FoV 		& $74'' \times 64''$ (hexagonal packed) \\
spatial sampling	& $2''.68$ per fiber diameter \\
fiber pitch		& $3''.5$ fiber-to-fiber \\
IFU-filter		& 2~inch / 50~mm round filter \\
PPCU-filter		& 1~inch / 25~mm round filter \\
wavelength range	& 400--900~nm (`dry' fibers) \\
\tableline %\tableline
{\bf PMAS spectrograph:} 	& \\
\tableline
PPak fiber slit 	& 0.15 $\times$ 94 mm, 382 fibers \\
slit-filter		& 140 $\times$ 35~mm filter \\
collimator 		& fully refractive 450 mm, F/3  \\
camera 			& fully refractive 270 mm, F/1.5  \\
reflective gratings	& 1200, 600 and 300 l/mm \\
detector 		& SITe 2k $\times$ 4k, 15 $\mu m$ pixels \\
linear dispersion	& 0.53, 1.2, 2.6 \AA /pixel (m=1)  \\
resolution		& 0.3 \AA /pixel, R$\approx$8000 (m=2)  \\
wavelength coverage 	& 600-3400 \AA /frame (1st order) \\
			& 400 \AA /frame (2nd order)\\
\tableline %\tableline
{\bf PMAS A\&G camera:} 	& \\
\tableline
A\&G FoV 		& 3.4 $\times$ 3.4 arcminutes  \\
A\&G plate scale 	& 0.2 arcseconds/pixel \\
A\&G-filter		& 4 $\times$ 2~inch/50~mm round filters \\
%filter options		& see {\tt www.caha.es/CAHA/Instruments/filterlist.html} \\
\tableline
\end{tabular}
\end{center}
\label{tab:pmas_param}
\end{table}

\vspace{5mm}
%%%%%%%%%%%%%%%%%%%%%%%%%%%%%%%%%%%%%%%%%%%%%%%%%%%%%%%%%%%%%
\section*{ACKNOWLEDGMENTS}

The authors would like to thank Ute Tripphahn \& the mechanical workshop
for the realization of various components of the PPak-IFU,
Thomas Hahn and Thomas Fechner (all AIP) for their help during commissioning.
MV likes to thank Matthew Bershady for valuable discussions
regarding design and construction issues. PPak was developed within
the \mbox{ULTROS} project, which is funded by the German ministry of
education \& research (BMBF) through Verbundforschungs grant 05-
\mbox{AE2BAA/4}.
AK, MR and MV gratefully acknowledge travel support from
the Deutsche Forschungsgemeinschaft (DFG).  We are thankful for the
excellent support we received from Calar Alto staff, in particular
Nicol\`as Cardiel, during the commissioning
and the subsequent observing runs with PPak. Special thanks to David
Lee (former AAO) for insights into aspects of the SPIRAL-B
manufacture.

\vspace{-5mm}

%%%%%%%%%%%%%%%%%%%%%%%%%%%%%%%%%%%%%%%%%%%%%%%%%%%%%%%%%%%%%
%%%%% References %%%%%
%%-----------------------------------------------------------
%\section*{References}

\clearpage

%%%%%%%%%%%%%%%%%%%%%%%%%%%%%%%%%%%%%%%%%%%%%%%%%%%%%%%%%%%%%

\end{document}